\documentclass[usenatbib,usedcolumn]{mn2e} 
\usepackage{epsfig}
\usepackage{subfigure}  
\usepackage{lscape}  
\usepackage{longtable} 
\title[Entropy and similarity in galaxy systems]{The Birmingham-CfA 
cluster scaling project - III: entropy and similarity in galaxy systems}
\author[T. J. Ponman, A. J. R. Sanderson and A. Finoguenov]{T. J.
  Ponman$^{1}$\thanks{E-mail: tjp@star.sr.bham.ac.uk}, A. J. R. 
  Sanderson$^{1,2}$ and A. Finoguenov$^{3}$\\
  $^{1}$School of Physics and Astronomy, University of
  Birmingham, Edgbaston, Birmingham B15 2TT, UK \\
  $^{2}$ Department of Astronomy, University of Illinois, 
  1002 West Green Street. Urbana, IL 61801, USA \\
  $^{3}$Max-Planck-Institut f\"ur extraterrestrische Physik, 
  Giessenbachstrasse, 85748 Garching, Germany \\
  }
 \date{Accepted 2002 ??.
       Received 2002 ??;
       in original form 2002 ??}

\pagerange{\pageref{firstpage}--\pageref{lastpage}}
\pubyear{2002}

\defcitealias{san03}{Paper~I}
\defcitealias{san03b}{Paper~II}
\defcitealias{pon03}{Paper~III}

\newcommand{\rmsub}[2]{\ensuremath{#1_{\mathrm{#2}}}} 
\newcommand{\ASCA}{\emph{ASCA}}
\newcommand{\cf}{{\textrm c.f.}}
\newcommand{\Chandra}{\emph{Chandra}}

\newcommand{\cm}{\ensuremath{\mbox{~cm}}}
\newcommand{\cmsq}{\ensuremath{\cm^{2}}}
\newcommand{\eg}{{\textrm e.g.}}

\newcommand{\etal}{{\textrm et al.\thinspace}}

\def\h70{\rmsub{h}{70}} 
\newcommand{\ie}{{\textrm i.e.}}
\newcommand{\keV}{\ensuremath{\mbox{~keV}}}

\newcommand{\Msol}{\rmsub{M}{\odot}}

\def\R200{\rmsub{R}{200}} 

\newcommand{\ROSAT}{\emph{ROSAT}}

\newcommand{\XMM}{\emph{XMM-Newton}}

\begin{document}

\maketitle

\label{firstpage}

\begin{abstract}
  \noindent

We examine profiles and scaling properties of the entropy of the
intergalactic gas in a sample of 66 virialized systems, ranging in
mass from single elliptical galaxies to rich clusters, for which
we have resolved X-ray temperature profiles. Some of the properties
we derive appear to be inconsistent with any of the models put
forward to explain the breaking of self-similarity in the baryon content
of clusters. In particular, the entropy profiles, scaled to
the virial radius, are broadly similar in form across the sample, apart
from a normalization factor which differs from the simple
self-similar scaling with temperature. Low mass systems do not
show the large isentropic cores predicted by preheating models,
and the high entropy excesses reported at large radii in groups
by \citet{fin02} are confirmed, and found to extend even to moderately
rich clusters. We discuss the implications of these results for
the evolutionary history of the hot gas in clusters, and suggest
that preheating may affect the entropy of intracluster gas primarily
by reducing the density of material accreting into groups and clusters
along cosmic filaments.
\end{abstract}


\section{Introduction}
\label{sec:intro}
In the widely accepted standard model for cosmic structure formation,
the Universe evolves hierarchically, as primordial density
fluctuations, amplified by gravity, collapse and merge to form
progressively larger systems. This hierarchical development leads to
the prediction of self-similar scalings between systems of different
masses and at different epochs. These scalings are also seen in
cosmological simulations involving only gravitationally driven
evolution, including compression and shock heating of the baryonic
matter. Such simulations \citep[\eg][]{nav95,fre99} result in haloes
in which the density profiles of both dark matter and baryonic material,
when radially scaled to the virial radius (which we define here to
be the radius \R200, within which the mean density of a system
is 200 times the critical density of the Universe) are almost identical
in virialized systems covering a wide range of masses, from
individual galaxies to rich clusters.

Given self-similar scalings of gas temperature and density,
scaled X-ray surface brightness profiles are also expected to be
similar. Furthermore, a simple scaling is expected between X-ray
luminosity, $L_X$, and temperature $T$. Assuming that the emission is
dominated by bremsstrahlung, $L_X \propto M_{\rm gas}^2 \, R^{-3} \,
T^{1/2}$, or $L_X \propto f_{gas}^2 \, T^{2}$, where $M_{gas}$ is the gas
mass within radius $R$, and
$f_{gas}=M_{gas}/M$ is the gas mass fraction. X-ray properties of
clusters deviate substantially from this simple scaling, and the
observed $L_X$:$T$ relation \citep{whi97,mar98} is considerably steeper than
$T^2$ in the cluster regime, and steepens further \citep{hel00b} in
galaxy groups. \citet{pon99} showed that the latter effect is due to the
suppression of $L_X$ in galaxy groups, arising from a reduction in
gas density in the inner regions of poor systems, relative to
richer clusters. 

It is instructive to view this in terms of the entropy of the
intergalactic medium (IGM), which in the self-similar case should
increase in a very simple scaling with the mean temperature of
virialized systems. In practice, it is found \citep{pon99,llo00} that
an excess in the entropy, above the self-similar prediction, is
apparent in the inner regions of poor clusters and groups, outside the
dense central regions in which cooling is expected to radiate away
entropy within the age of the Universe. This effect has been referred
to as the `entropy floor', with the implication that additional
physical processes, beyond gravity and resulting compression and shock
heating, have acted to set a lower limit to the entropy which the gas
in collapsed haloes can have.

A great deal of theoretical work has been devoted to explaining this
phenomenon over the past few years. As we will discuss in some detail
later, the explanations proposed fall into three main classes:
the gas has been heated either at an early epoch, before clusters
were assembled \citep{kai91,evr91,cav97,bal99,val99,toz01}, or it has
been heated {\it in situ} by star formation and/or energy input from
active galactic nuclei (AGN) \citep{bow97,loe00,voi01,nat02}. Alternatively,
some authors have argued \citep{kni97,bry00,pea00,mua01,wuX02,dave02} that
cooling alone will remove low entropy gas from the centres of haloes,
producing a very similar effect to non-gravitional heating.

In the present paper we aim to confront these models with the
observed properties of the hot gas in a large sample of galaxy systems,
spanning a wide range in total mass. In the fullness of time, high
quality X-ray observations of the density and temperature structure
of the IGM will be available from \XMM\ and \Chandra. However, at present,
such observations are sparse, and it is essential to have a broadly 
representative and wide-ranging sample of virialized systems in order
to study scaling properties. The value of this in the context of
similarity-breaking in clusters has already been shown by a number 
of earlier studies. 

In the present, and companion papers, \citet{san03} (Paper~I) and
\citet{san03b} (Paper~II), we examine scaling properties derived from
the largest sample of virialized systems with resolved X-ray temperature
profiles yet assembled. Following a brief description of the sample
and our analysis in section~\ref{sec:sample} (details are given in
Paper~I), we present the profiles and scaling properties for the
entropy and temperature across our sample in 
sections~\ref{sec:distributions} and \ref{sec:scaling}. These results
are used in section~\ref{sec:discuss}, along with relevant results
from Papers~I and II, to test the various models proposed to account
for the entropy floor, and finally in section~\ref{sec:conc} we draw
our conclusions from this study, and propose a new model to explain
the behaviour of the entropy of intracluster gas.

\section{Sample and analysis}
\label{sec:sample}
Our sample comprises 66 virialized systems, from rich clusters of galaxies,
through groups and down to the level of individual galaxy-sized haloes. In
\citet[][hereafter \citetalias{san03}]{san03}, we reported a detailed
study of the 3-dimensional X-ray properties of this sample, based on
data from the \ROSAT\ and \ASCA\ observatories, which we
assembled from the work of three separate investigators \citep{mar98b,
mar98,mar99,mar97,mar96,fin99,fin00,fin00b,fin01b,llo00},
combined with a number of cool groups analysed specially to
provide better coverage of the crucial low end of the mass range. 
To each system we fitted analytical functions, describing the gas
density and temperature variation with radius, outside any central
cooling region, which was excised or fitted separately. This approach 
allows us to put the X-ray data from the three earlier studies
on a unified footing, and gives us the freedom to
extrapolate the gas properties to arbitrary radius. We used a beta model
to parametrize the density and specified the temperature variation with
either a linear ramp or a polytropic IGM. We have used these data
to determine the gravitating mass profile and thus to calculate radii of
overdensity in a self-consistent manner. Similarly, we have derived mean
temperatures for each system, by averaging the gas temperature within
0.3\R200, weighted by its luminosity \citepalias{san03}.

\section{Entropy and temperature distributions}
\label{sec:distributions}
For convenience, we define `entropy' in the present paper as
\begin{equation}
S=T/n_e^{2/3}  \keV\cmsq ,
\end{equation}
which relates directly to observations. This has been referred to by
a number of authors as the `adiabat', since (apart from a constant
relating to mean particle mass) it is the coefficient relating pressure
and density in the adiabatic relationship $P=K \rho^\gamma$. Hence $S$
is conserved in any adiabatic process.
Note that the true thermodynamic entropy is related to our definition
via a logarithm and additive constant.

\begin{figure*}
  \centering
  \subfigure{
    \includegraphics[angle=270,width=8.6cm]{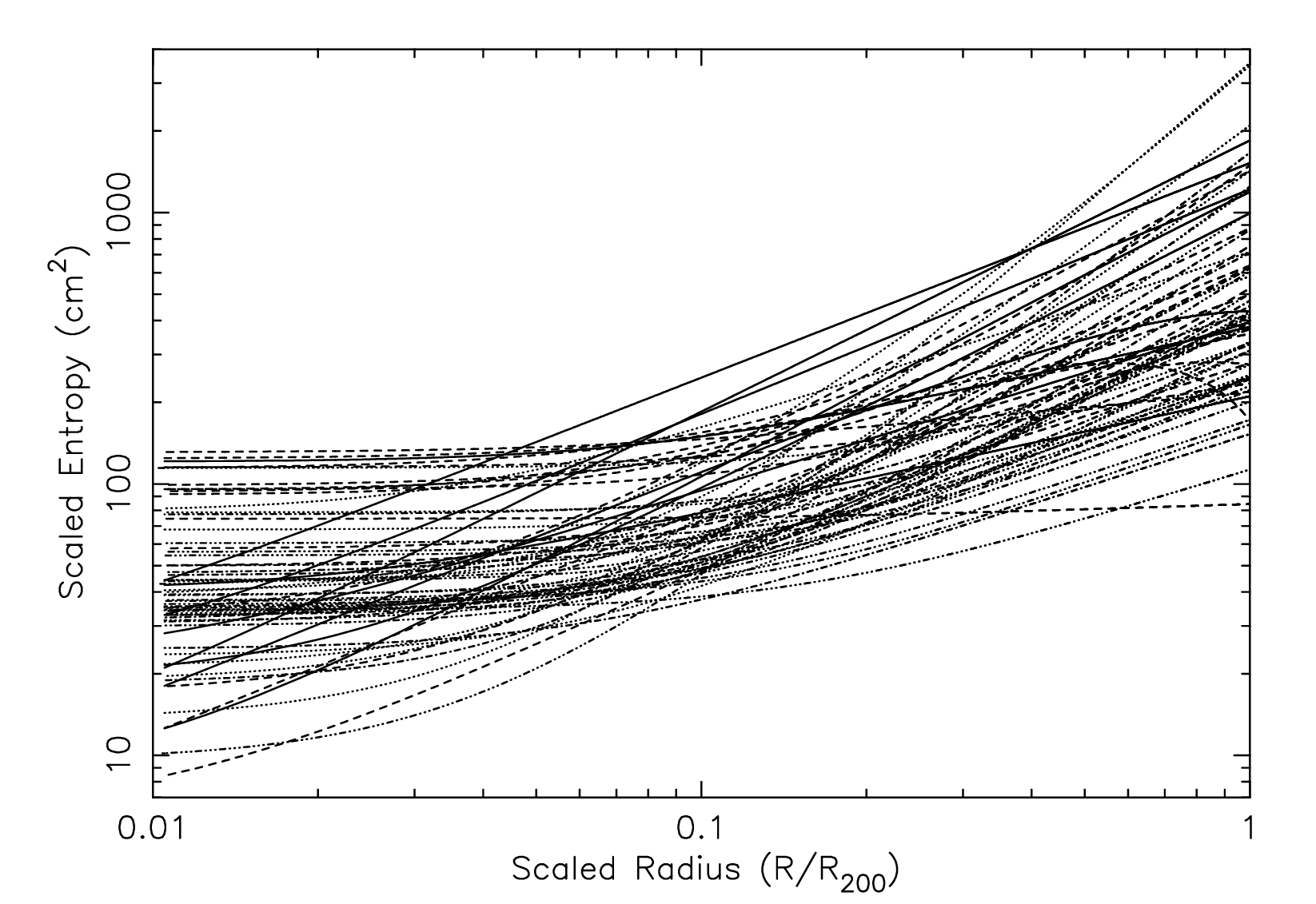}}
  \hspace{-0.3cm}
  \subfigure{
    \includegraphics[angle=270,width=8.6cm]{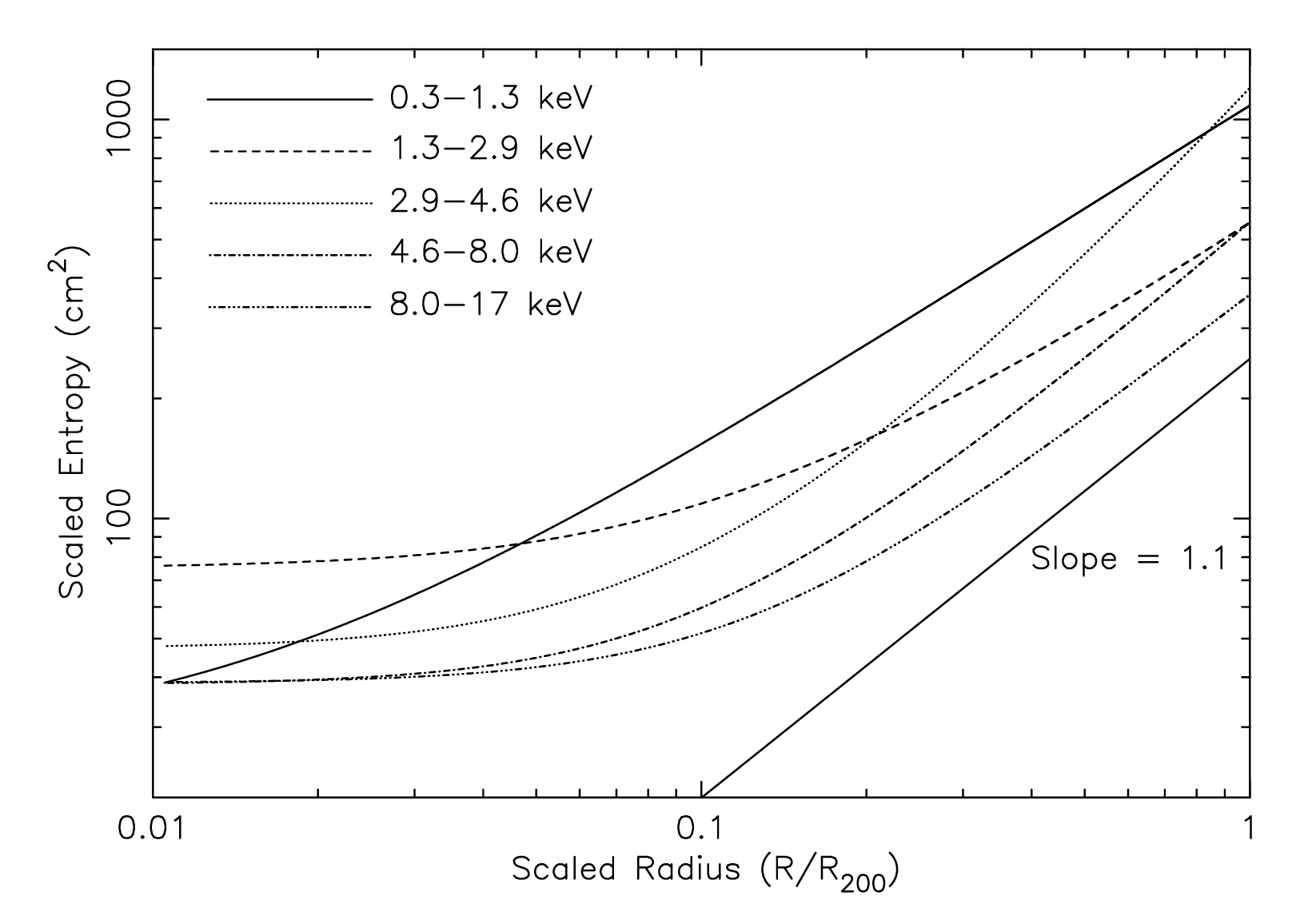}}
  \caption{Entropy profiles for each system in our sample (left panel)
    derived from our fitted models, each scaled
    by $1/T$. The line style of each profile denotes the mean
    temperature of the system, as described below.
    In the right panel, these profiles (excluding the two galaxies, as
    discussed in the text) have been grouped into 
    temperature bands: 0.3--1.3\keV\ (solid), 1.3--2.9\keV\ (dashed),
    2.9--4.6\keV\ (dotted), 4.6--8\keV\ (dot-dash) and 8-17\keV\ 
    (dot-dot-dot-dash). The bottom line shows the slope of 1.1 expected 
    from shock heating. Its normalization is arbitrary.}
  \label{fig:s_r}   
\end{figure*}

In Fig.~\ref{fig:s_r}, we overlay the scaled entropy profiles for
all 66 systems in our sample. Under the assumption that all these
systems form at the same redshift, their mean mass densities
should be identical. Hence in the simple self-similar case, where
all have similar profiles and identical gas fractions, $S$ will simply
scale with mean system temperature $T$. We apply this scaling, and
scale the radial coordinate to $R_{200}$ for each system, derived from 
our fitted models (Paper I). It can be seen that the entropy profiles
of the cooler systems, scaled in this way, tend to be significantly
higher than those of richer clusters.
To see this more clearly, in the right panel of Fig.~\ref{fig:s_r}, we
show the profiles grouped into bands of mean system temperature.
In this grouped plot we have excluded the two galaxies in our sample, 
whose profiles may be dominated by stellar wind losses, rather than 
by the processes operating in groups and clusters. 

It can be seen that there is a strong tendency for the scaled entropy
to be higher, at a given scaled radius, in cooler systems. Simulations
and analytical models of cluster formation involving only gravity and 
shock heating, produce entropy profiles with logarithmic slopes of 
approximately 1.1 \citep{toz01}, which agrees rather well with the
slope of our profiles outside 0.1\R200. The mean profile in the 
2.9--4.6\keV\ band is significantly affected by two systems, visible
in the left hand panel, which have strongly rising profiles.
One of these is AWM7, for which the temperature profile is subject
to especially large systematic uncertainties, as a result of the
strong cooling flow, as discussed in Paper I. The temperature model
adopted, from the analysis of \citet{llo00}, rises strongly with radius.
As a result, the slope of $S(r)$ in this band is almost certainly
overestimated in the figure.

One possibility which we can discard right away, is that the observed
scaling results from systematic differences in the formation epochs of
low and high mass systems. In hierarchical models, low mass systems
are expected to form, on average, earlier than high mass ones. Hence
they should tend to have higher mean densities, and therefore {\it
lower} gas entropy -- the opposite of what we observe.  It is true
that, due to the usual selection effects, the rarer and more luminous
massive systems in our sample tend to be at higher redshifts than the
groups. Hence if, contrary to expectations, all systems virialized at
the redshift we observe them, then the more massive systems would be
more dense, and have lower entropies. However, this effect is
considerably smaller than the trend which we observe. The highest
redshift clusters in our sample are at $z\sim 0.2$, hence their mean
densities, scaling as $(1+z_f)^3$, could be 70\% higher than the
lowest redshift systems, and their entropies (scaling as $n^{2/3}$)
consequently lower by a factor of 1.44. All but four members of our
sample lie at $z<0.1$, and hence would have entropies changed by less
than 20\% as a result of such a redshift-dependent density scaling. In
contrast, Fig.~\ref{fig:s_r} shows that the scaled entropy actually
differs by a factor ~3 between the high and low temperature bins for
our sample.

At small radii, our fitted models exclude the effects of
cooled gas, since any central cooling region is excised, or represented
by a separate component in our analysis (sect.~\ref{sec:sample}).
This central cooling region is present in 54 of our 66 systems
(see Table 1 in Paper~I), and in these cases its
radius ($r_{\rm cool}$) ranges from 0.03$R_{200}$ to 0.2$R_{200}$,
with a median value of $r_{\rm cool}=0.06R_{200}$.
Apart from the coolest systems, our entropy profiles
generally flatten inside $0.1\R200$.  This effect is also seen in many
cosmological simulations \citep{fre99} even in the absence of
non-gravitational heating and cooling processes, and appears to result
from the introduction of a core into the gas density distribution, due
to transfer of energy between baryonic and dark matter during merger
events \citep{eke98}.  Preheating models generally predict large
isentropic central regions in low mass systems, which are not seen in
our data. We will return to this point in section~\ref{sec:discuss}
below.

\begin{figure*}
\includegraphics[angle=270,width=13cm]{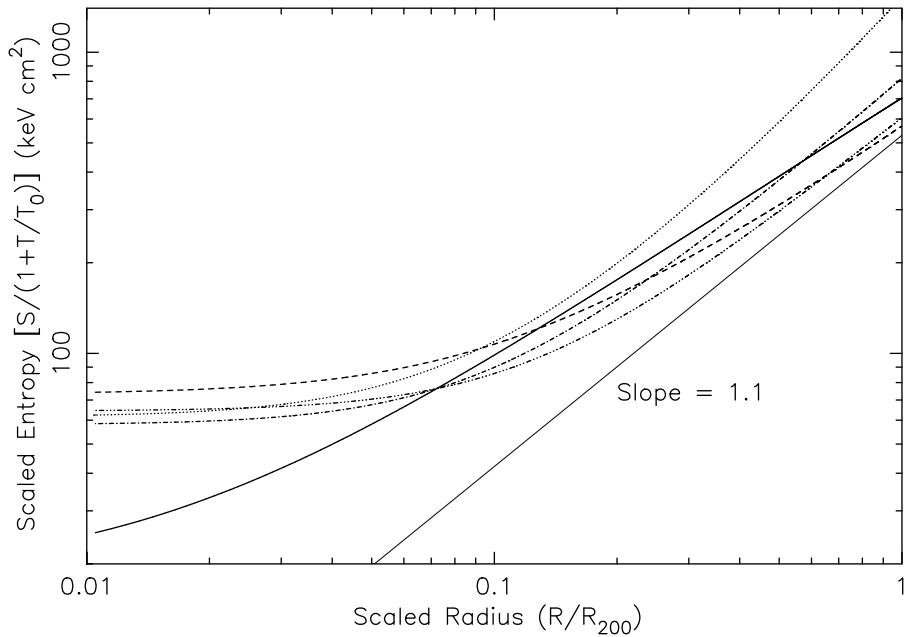}
\caption{ \label{fig:s_rDS}
  The variation of gas entropy (scaled by $(1+T/T_0)^{-1}$) with scaled
  radius, grouped by system temperature. The solid line represents the
  coolest systems (excluding the two galaxies) (0.3--1.3\keV), increasing
  in temperature through dashed (1.3--2.9\keV), dotted (2.9--4.6\keV), 
  dot-dashed (4.6--8\keV) and finally dot-dot-dot-dashed
  (8--17\keV). The lower solid line (with arbitrary normalization)
  indicates the slope of 1.1 expected from shock heating.}
\end{figure*}

The preheating model of \citet{dos02} predicts that entropy should scale
according to $(1+T/T_0)$, rather than $T$, where $T_0$ is a constant
related to the degree of preheating, which they estimate as
$T_0$=2\keV\, to provide a best fit to the entropy floor data of
\citet{llo00}. In Fig.~\ref{fig:s_rDS}, we show the effect of this scaling
on our temperature-grouped entropy profiles. This scaling does indeed
bring the profiles into good agreement, apart from the fact
that our coolest systems show little sign of any central entropy
core. Note from the left hand panel in Fig.~\ref{fig:s_r}, that this
is a general feature of almost all the entropy profiles for cool
systems, rather than a result of averaging together systems with cores
and others with strongly dropping central entropies.

\begin{figure*}
\includegraphics[angle=270,width=13cm]{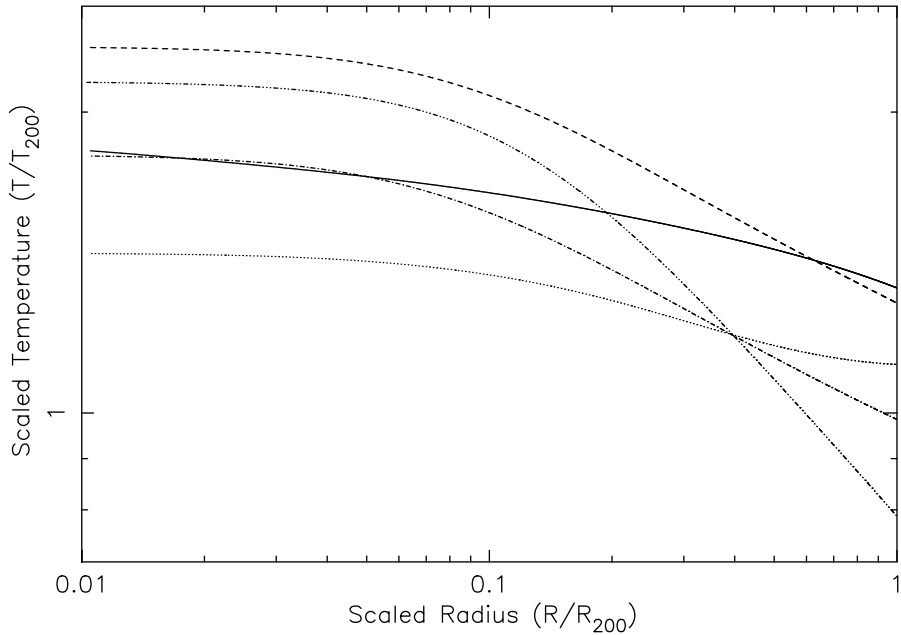}
\caption{ \label{fig:t_r_grp}
  The variation of gas temperature (scaled by $T_{200}$) with scaled
  radius, grouped by system temperature. The solid line represents the
  coolest systems (excluding the two galaxies) (0.3--1.3\keV), increasing
  in temperature through dashed (1.3--2.9\keV), dotted (2.9--4.6\keV,
  excluding AWM~7), dot-dashed (4.6--8\keV) and finally dot-dot-dot-dashed
  (8--17\keV).}
\end{figure*}

It is also instructive to compare temperature profiles across the
range of masses in our sample. We use the virial radii and 
masses calculated from our fitted models to compute the
virial temperature
\begin{equation}
T_{200} = \frac{G M \mu m_p}{2 R_{200}} ,
\end{equation}
expressed in \keV. This is used to normalize each temperature
profile, and the scaled profiles are grouped in temperature bands, to reduce
the large scatter and make trends more obvious. The result is shown
in Fig.~\ref{fig:t_r_grp}. As with the grouped
entropy plots, we have excluded the two galaxies
from these profiles. We have also omitted AWM7 from the 2.9--4.6\keV\
band, since its highly uncertain (see discussion above) and strongly 
rising $T(r)$ profile distorts the mean profile for the whole band.

Raising the entropy of the IGM results in
increased temperatures as well as lower densities, and 
for a given level of entropy increase this effect
will be most prominent in low mass systems, where the 
`natural' shock-generated entropy is lower. As can be seen, our
observations do indeed show such an effect at large radii -- at 
\R200\ the scaled temperatures are ordered in precisely this way.
However, at smaller radii, this is not the case, and in particular,
the coolest systems display very flat temperature profiles, whilst
the most massive clusters show the largest rise in temperature
between \R200\ and the centre. This behaviour is not what is predicted
by most models, as we will discuss later.

\section{Scaling properties}
\label{sec:scaling}
The claim of the discovery of an entropy floor in galaxy systems
\citep{pon99} was based on the measurement of gas entropy at a scaled
radius of $0.1$\R200, in systems spanning a wide temperature range.
This radius was chosen to lie close to the centre, where shock-generated
entropy should be a minimum, hence maximising the sensitivity to any
additional entropy, whilst lying outside the region where the cooling time
is less than the age of the Universe, and hence the entropy may be reduced.
This initial study was improved by \citet{llo00}, who avoided the isothermal
assumption made by \citeauthor{pon99}, and derived an entropy floor
value of $139\,h_{50}^{-1/3}$\keV\cmsq\ from a sample of 20 systems, 
which is essentially a subset of the present work.

\begin{figure*}
\includegraphics[angle=270,width=13.5cm]{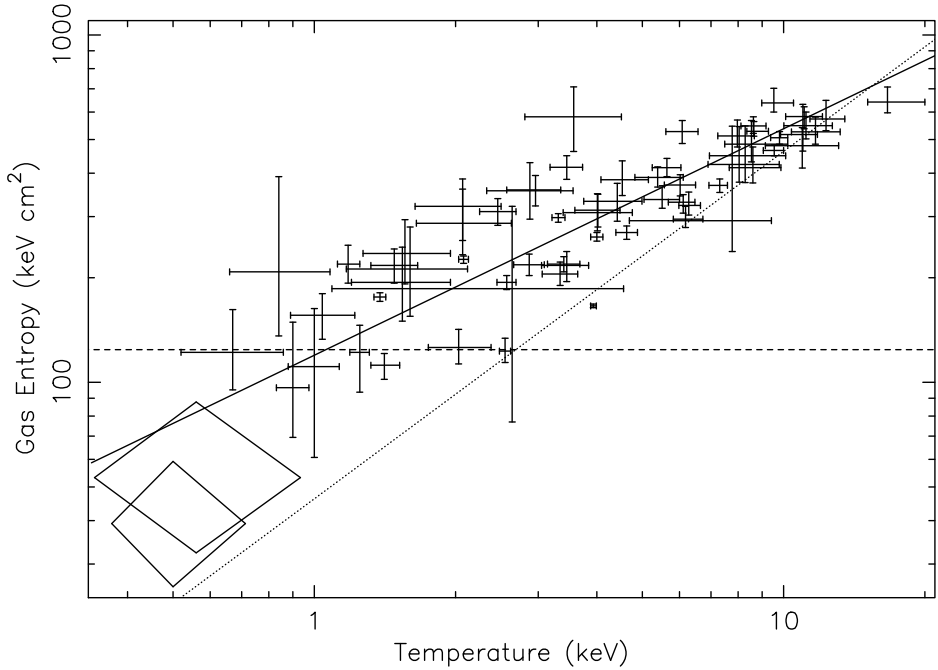}
\caption{ \label{fig:S0.1_T}
  Gas entropy at 0.1\R200\ as a function of system temperature. The
  diamonds represent the two galaxies. The solid line is the best fit to
  the barred points. The 
  dotted line has a self-similar slope of 1 and is normalized to the mean
  of the hottest 8 clusters. The dashed line is the entropy floor of 
  124\,\h70$^{-1/3}$\keV\cmsq\ from \citet{llo00}.}
\end{figure*}

In Fig.~\ref{fig:S0.1_T}, we show the corresponding plot from our much
larger sample. With the benefit of this increase in sample size, the
trend looks rather different from its appearance in \citet{llo00},
where a dearth of systems in the 1.5--3.5\keV\ band led to the
interpretation of a relation which followed the self-similar line down
from hot systems, flattening towards a floor value of
$139\,h_{50}^{-1/3}$\keV\cmsq, corresponding, with our value of $H_0$,
to 124\,\h70$^{-1/3}$\keV\cmsq.  However, it appears from our data
that the behaviour is rather that of a slope in $S(T)$ which is
significantly shallower than the self-similar relation $S\propto T$
throughout.  Using unweighted orthogonal regression \citep{iso90}, 
we obtain a
relation $S(0.1\R200)\propto T^{0.65\pm0.05}$, which is marked in
Fig.~\ref{fig:S0.1_T}. The \citeauthor{llo00} floor value lies at the
bottom of this trend, but it is not clear whether it sets a lower
limit to the entropy in galaxy groups, since no groups with $T\la
0.6$\keV\ have been bright enough for detailed study to date. The two
galaxies in our sample do clearly have entropies which lie below the
floor level, however much or all of the hot gas in these systems may
have its origin in stellar mass loss, rather than retained primordial
material \citep{osu01a}, so it may well be fortuitous that they fall
close to the trend set by the groups and clusters.

\begin{figure*}
\includegraphics[angle=270,width=13.5cm]{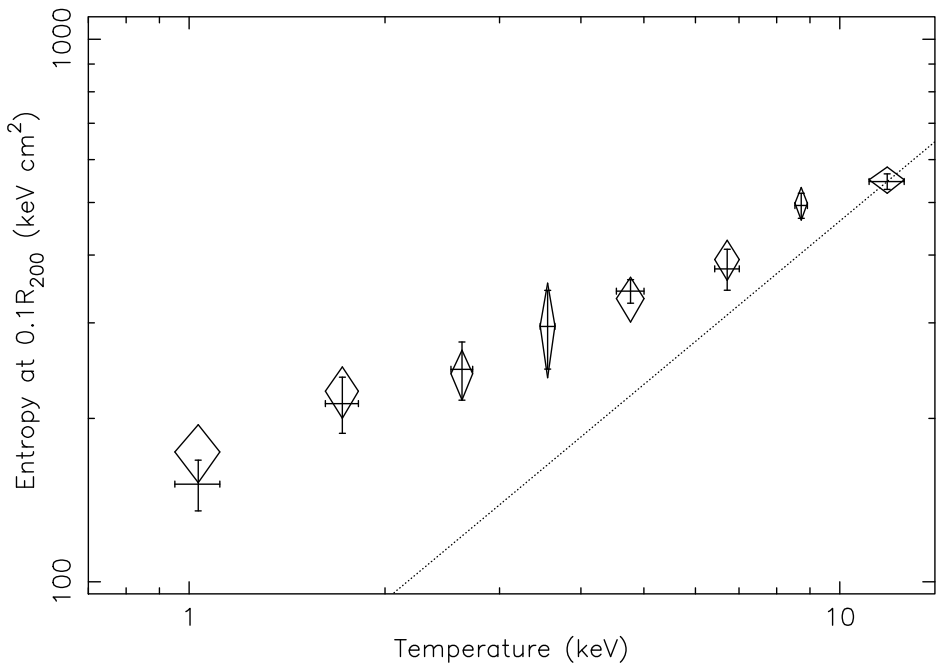}
\caption{ 
  Gas entropy at 0.1\R200\ as a function of system temperature, excluding
  the two galaxies and grouped to a minimum of 8 points per bin.
  Barred crosses are based our measured values of \R200, whilst
  diamonds result from calculation of \R200\ from mean system temperature
  using the $T^{1/2}$ scaling of \citet{nav95}. The dotted line shows the
  self-similar slope of 1, normalized to the 8 hottest clusters.}
\label{fig:S0.1_T_grp}
\end{figure*}

Grouping the points into temperature bins (Fig.~\ref{fig:S0.1_T_grp})
makes the unbroken nature of the trend very clear. An unweighted
orthogonal fit to these grouped points gives a logarithmic 
slope of $0.57\pm0.04$, slightly flatter than that derived from
Fig.~\ref{fig:S0.1_T}. Note that these results suggest the presence of
excess entropy of $\sim 100$\keV\cmsq\ in {\it all} temperature bands,
relative to the hottest systems.  Fig.~\ref{fig:S0.1_T_grp} also
compares the effect on the relation of calculating \R200\ from a
measurement of the mass profile, compared to assuming a scaling $\R200
\propto T^{1/2}$, as was done (employing the formula derived from
simulations by \citet{nav95}) by \citet{pon99} and \citet{llo00}. As
discussed in Paper~I, the \citeauthor{nav95} formula agrees reasonably
well with our measurements in rich clusters, but in cool systems, the
upward bias in temperature, relative to simulations such as those of
\citet{nav95}, which include only gravitational physics, causes the
$\R200 \propto T^{1/2}$ scaling to overestimate \R200\, leading in
turn to an upward bias in the entropy (since $S$ rises with
radius). This may have contributed, in previous studies, to the
appearance of flattening in the $S(T)$ relation towards low $T$.

Whilst measurements close to the cluster centre provide the most
sensitive probe of excess entropy, detection of additional entropy at
large radii is especially interesting, since many preheating models
predict that the rise in entropy is essentially restricted to those
central regions where shock-generated entropy is less than the floor
value set by preheating. \citet{fin02} were the first to find evidence
for excess entropy at a much larger radius, \rmsub{R}{500},
corresponding (Paper II) to $\sim \frac{2}{3}$\R200. They argued that the high
excess entropy observed at large radii in groups and poor clusters
indicates that their IGM is dominated throughout by the effects of
preheating, with shocks playing little or no role. We aim, in this
paper, to check this result with our larger sample.

\begin{figure*}
  \centering
  \subfigure{
    \includegraphics[angle=270,width=9.0cm]{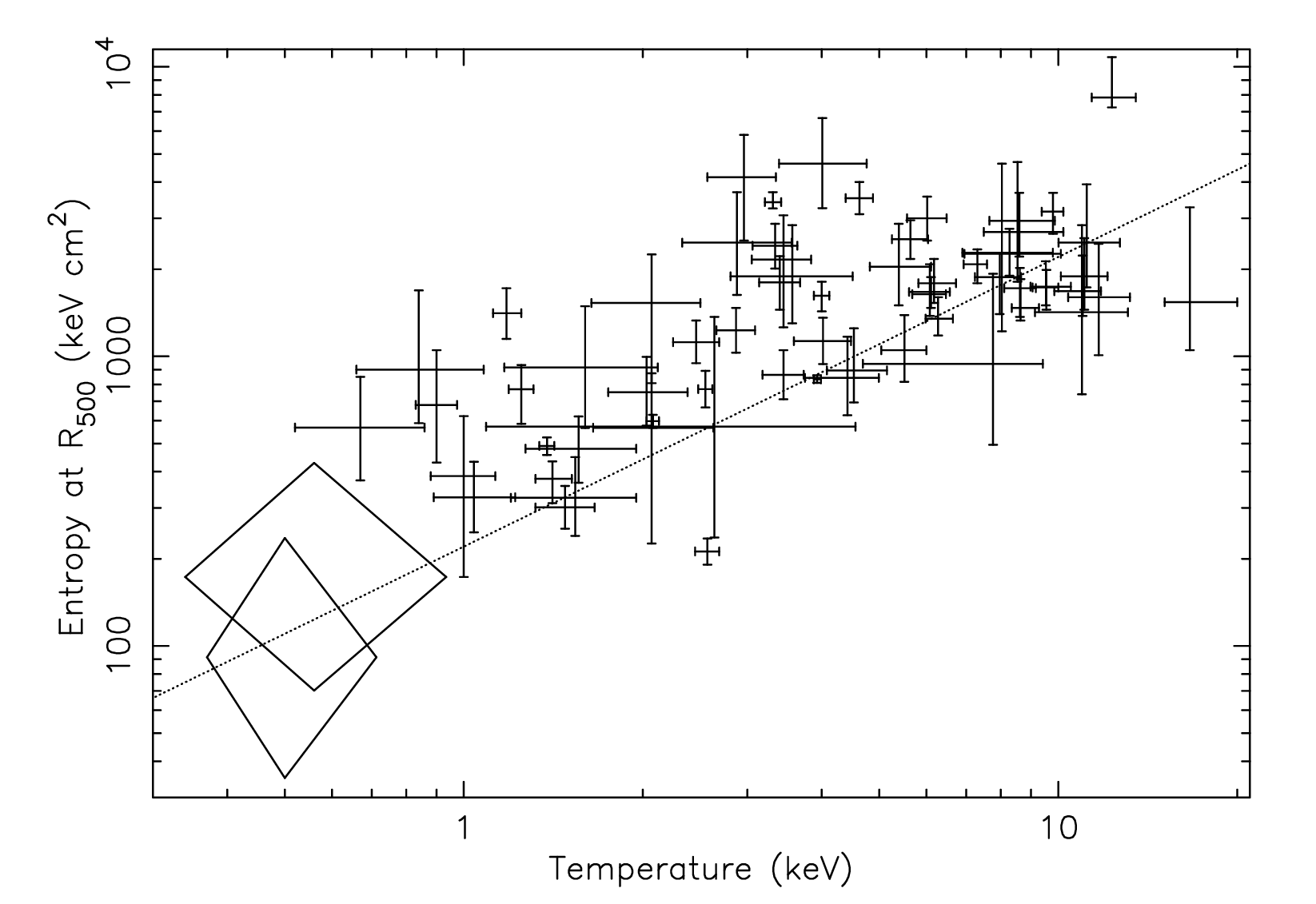}}
  \hspace{-0.7cm}
  \subfigure{
    \includegraphics[angle=270,width=9.0cm]{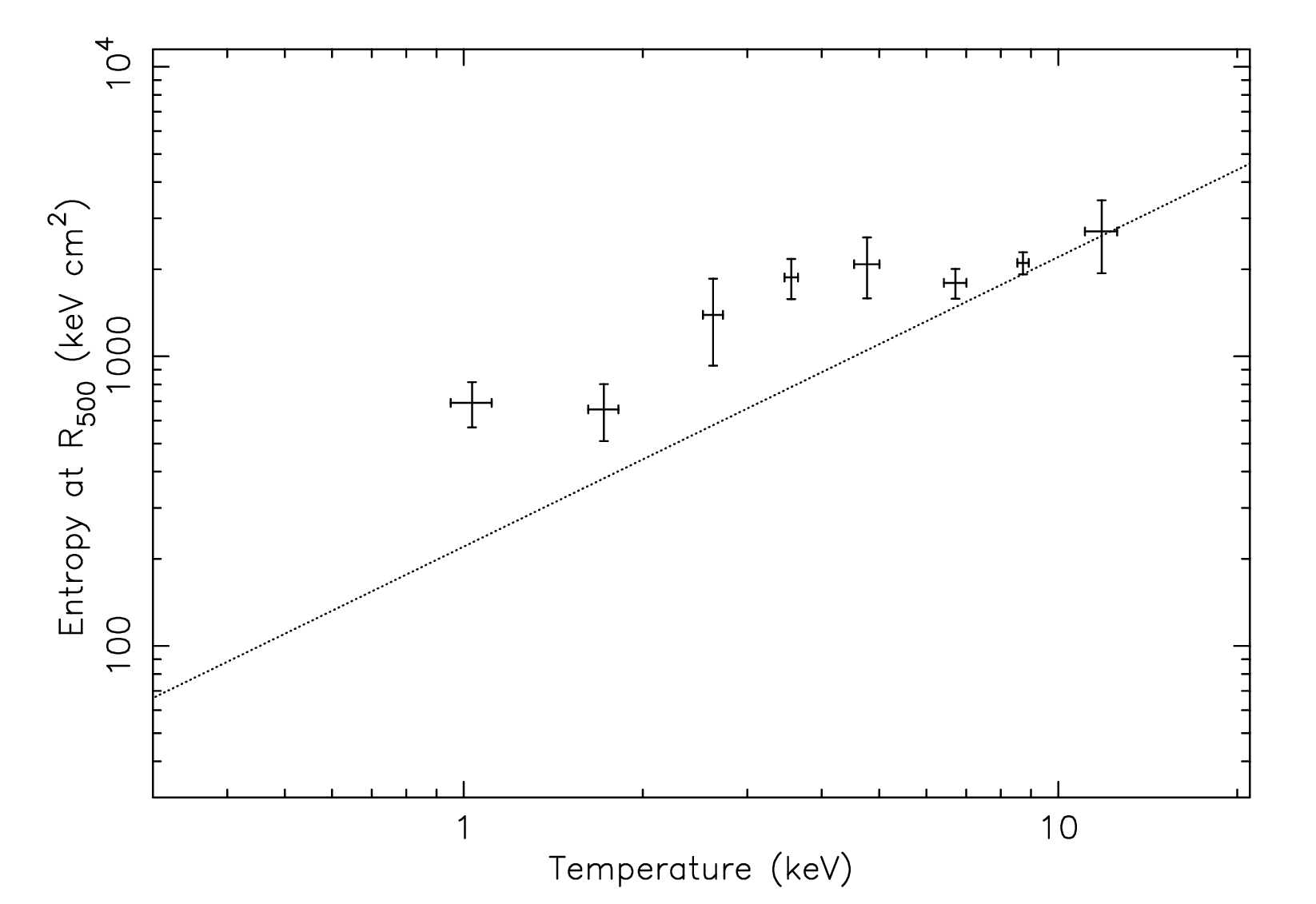}}
  \vspace{-0.3cm}
  \caption{Entropy at $R_{500}$ as a function of system temperature for
    each of the 66 systems, with the galaxies marked as diamonds (left panel)
    and grouped to a minimum of 8 points per bin, excluding the galaxies
    (right panel). The dotted line shows the self-similar slope of 1,
    normalized to the 8 hottest clusters.}
  \label{fig:S500_T}   
\end{figure*}

Our data, shown in Fig.~\ref{fig:S500_T}, confirm the existence of substantial 
excess entropy at \rmsub{R}{500}, above a self-similar extrapolation 
from the values seen in rich clusters (dotted line). The trend is more clearly
seen in temperature-grouped data shown in the right hand panel, and the
logarithmic slope of $S(T)$ is similar to that seen at 0.1\R200. The 
departure from self-similar entropy values is
apparent not only in the coolest systems, but also in quite rich
clusters. In fact, the largest absolute values of excess entropy
($\sim 1000$\keV\cmsq) are seen in clusters with $T\sim$3--4\keV.

\begin{figure*}
\includegraphics[angle=270,width=13.5cm]{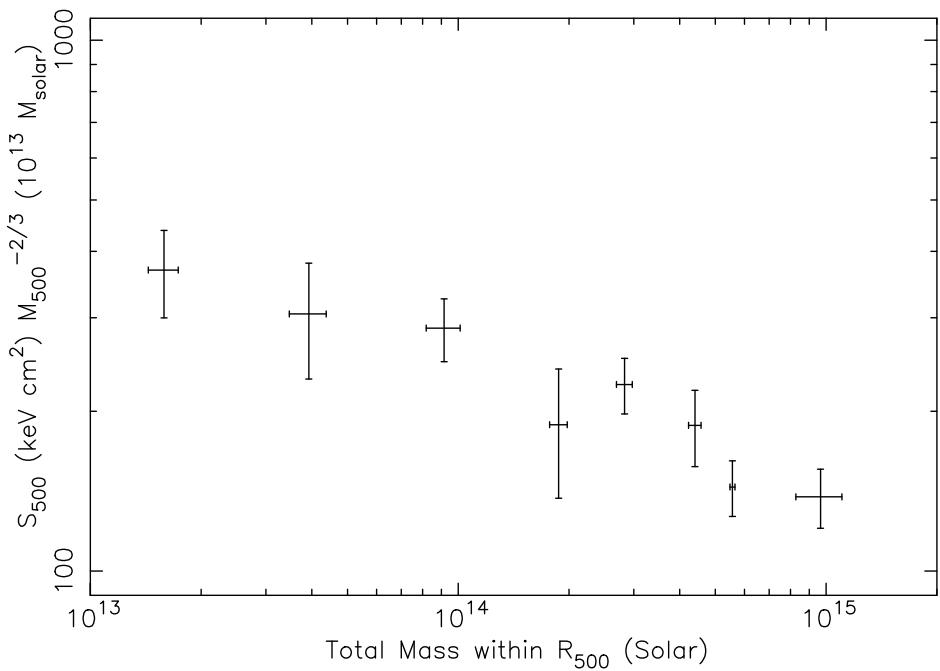}
\caption{
  Gas entropy at $R_{500}$, normalized by $M_{500}^{2/3}$, as a function of
  the total mass within $R_{500}$ (excluding the two galaxies) and grouped
  to a minimum of 8 points per bin. }
\label{fig:S500_Mnorm}
\end{figure*}

For direct comparison with \citet{fin02}, we show in
Fig.~\ref{fig:S500_Mnorm} the entropy at \rmsub{R}{500}\ scaled by
$M_{500}^{-2/3}$, and grouped into mass bins to suppress fluctuations.
For a set of self-similar systems, virialising at the same epoch (and hence 
having the same mean density), this scaling should renormalize the
entropy to a constant value, independent of system temperature (since
$S \propto T \propto M^{2/3}$).  Clearly real clusters do not follow
this self-similar law.  Whilst our plot is broadly consistent in shape
with Fig.~3 of \citet{fin02}, our larger sample again reveals
important additional features.  \citet{fin02} concluded that excess
entropy was only present in systems with $M_{500}\la
10^{14}\,h_{50}^{-1}$\Msol\ ($7\times 10^{13}$\Msol\ for our choice of
$H_0$), whereas it is clear from our data that a trend in scaled
entropy is present across the full mass range of clusters and groups.

\citet{fin02} also examined the entropy at a fixed value of enclosed
gravitating mass ($3\times10^{13}\,h_{50}^{-1}\Msol$) over their
sample of systems. The infall velocity of gas into an
accretion shock should be similar to the free fall velocity at the
shock radius, which depends upon the enclosed mass and mean
density of the Universe at the epoch in question.
In clusters, an enclosed mass of $3\times10^{13}\,h_{50}^{-1}\Msol$ lies
deep within the system, whilst in small groups it can lie close to \R200.
Since cluster cores are generally assembled at higher redshifts than
the outskirts of groups, one expects the gas density in the shell
under consideration to be higher, and the entropy generated by the thermalised
kinetic energy is therefore lower, by a factor ($1+z_f$), where $z_f$ is
the redshift at which the shell was accreted. In practice, \citet{fin02} found
the entropy of this shell to be bimodal across their sample, with a typical
value of $\sim$300\keV\cmsq\ for systems with $T\la 3$\keV, and with
considerably lower values (scattered over the range 100-300\keV\cmsq) in
hotter systems. Such a distribution cannot be accounted for on the basis
of a $1+z_f$ scaling, and \citeauthor{fin02} suggested instead that the entropy
in cool systems is set by preheating, taking place at $z\sim 3$, after
many cluster cores had already collapsed.

\begin{figure*}
  \centering
  \subfigure{
  \includegraphics[angle=270,width=9.0cm]{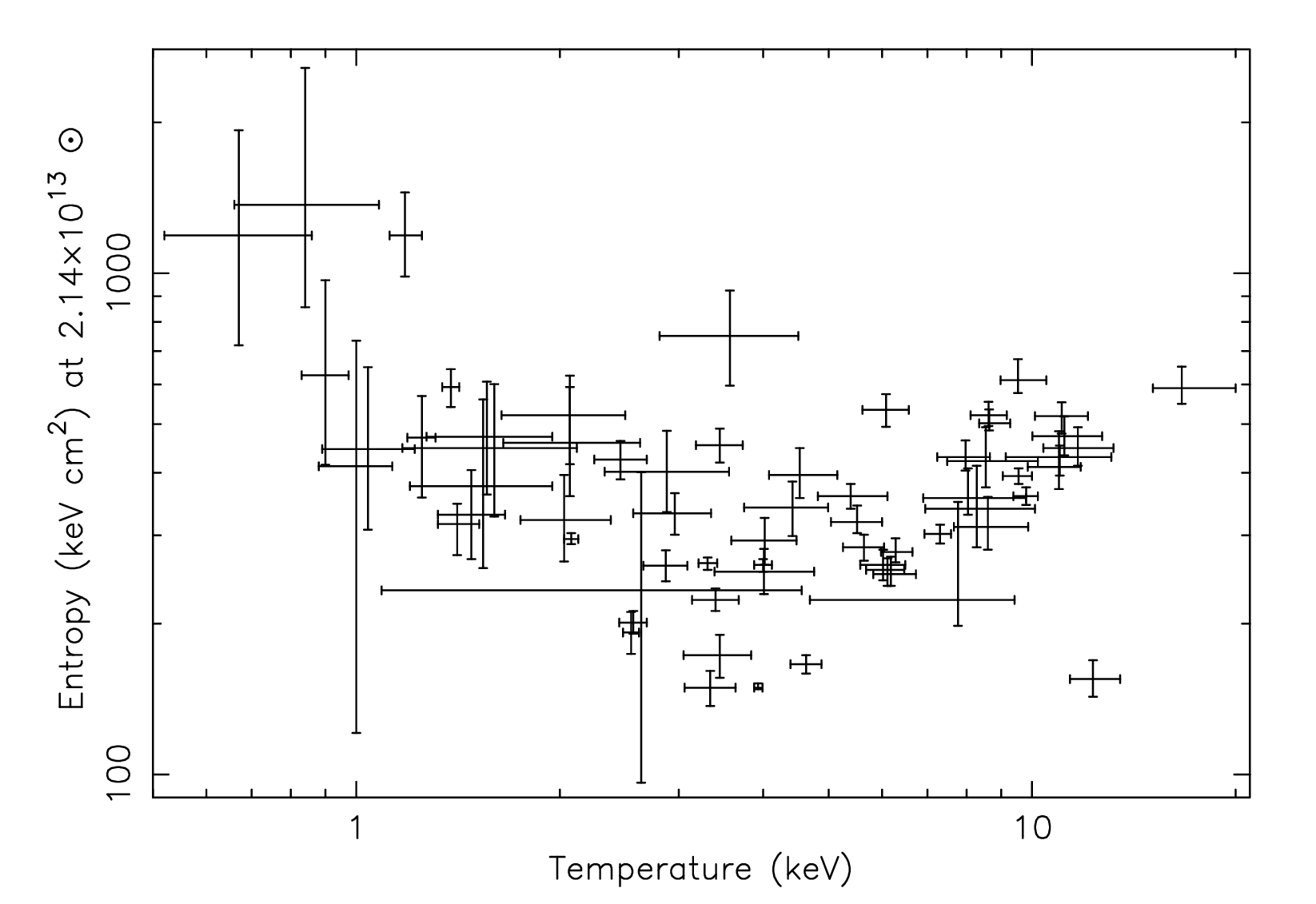}}
  \hspace{-0.7cm}
  \subfigure{
    \includegraphics[angle=270,width=9.0cm]{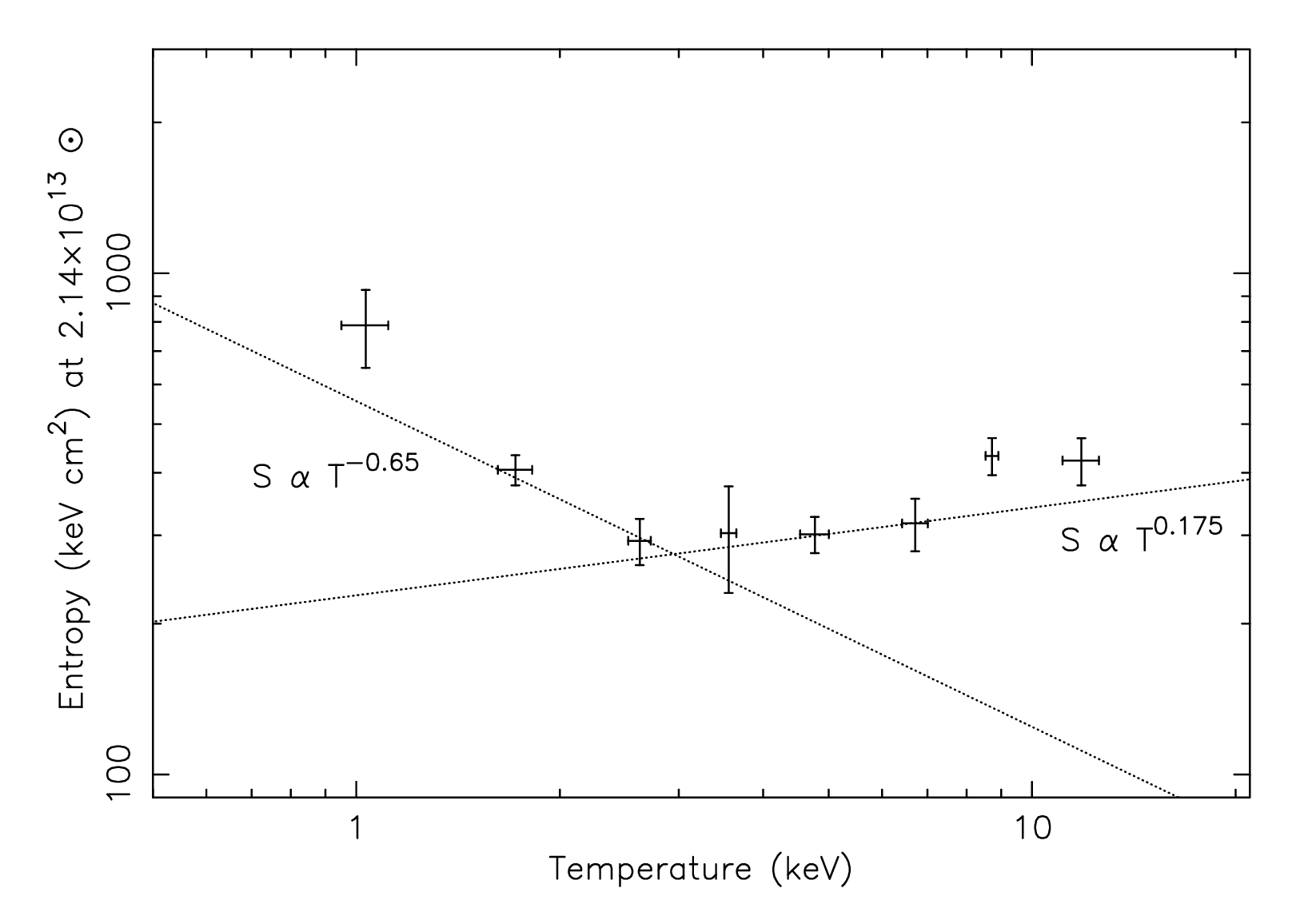}}
  \vspace{-0.3cm}
  \caption{ Left: entropy at the radius enclosing a mass of
  $3\times10^{13}\,h_{50}^{-1}\Msol$ as a function of system temperature
  (excluding the two galaxies). Right: the same data grouped into bins
   containing 8 points or more, with simple limiting trends, discussed
   in the text, marked for comparison. }
  \label{fig:SM3e13_T}   
\end{figure*}

The corresponding plot for our sample, which is almost double the size
of that of \citeauthor{fin02}, is shown in
Fig.~\ref{fig:SM3e13_T}. For direct comparison we have derived
entropies at an enclosed mass of $2.14\times10^{13}\,h_{70}^{-1}\Msol$,
allowing for our different choice of Hubble constant. 
Note that in some of the most massive systems, the radius enclosing 
$2.14\times10^{13}\,h_{70}^{-1}\Msol$ lies (just) within the
cooling region, and hence our derived entropy value is effectively
extrapolated, using the model for the non-cooling component (see section~3).
This affects six of the clusters which contribute to the highest
temperature bin in the right panel of Fig.~\ref{fig:SM3e13_T}, and a handful
of cooler clusters. Our results look
significantly different from those of \citet{fin02}. The entropy
appears to be non-monotonic, with a minimum value in systems with
$T\sim$3--4\keV. From a purely phenomenological perspective, this
non-monotonic behaviour can be understood in terms of mass profiles
which take the NFW form \citep{nav97}
\begin{equation}
\rho(r)=\frac{\rho_s}{(r/r_s)(1+r/r_s)^2} , 
\end{equation}
combined with entropy profiles which rise as $S\propto r^{1.1}$, outside 
some flatter central core. In groups, the mass shell under consideration here
lies well outside $r_s$, where enclosed mass grows roughly linearly with 
radius,
\begin{equation}
M(<r)\sim C r T_{200}^{3/2} , 
\end{equation}
so that the radius at which an enclosed
mass of $2.14\times10^{13}\,h_{70}^{-1}\Msol$ is achieved scales as
\begin{equation}
r_*\propto T_{200}^{-3/2}.
\end{equation}
Since the entropy at this radius scales as
\begin{equation}
S(r_*)\propto T_{200}r_*^{1.1} ,
\end{equation}
we expect 
\begin{equation}
S(r_*)\propto T_{200}^{-0.65},
\end{equation}
which gives a reasonable match to the slope at $T<3$\keV, as shown
in the right hand panel of Fig.~\ref{fig:SM3e13_T}.

In more massive systems ($M_{200}>2\times 10^{14}\,h_{70}^{-1}\Msol$,
$T>$3\keV), the mass shell we are studying moves inside the scale
radius $r_s$, so that the enclosed mass grows as $r^2$. An analysis
similar to that above then results in
\begin{equation}
S(r_*)\propto T_{200}^{0.175},
\end{equation}
provided that $S(r)\propto r^{1.1}$ at these small radii. However,
we saw in Fig.~\ref{fig:s_r} that for all but the coolest systems,
$S(r)$ flattens within the cluster core, in which case the positive
slope of the $S(r_*)$:$T$ relation will be steeper than 0.175, as is
observed.

The above arguments seem able to explain the general form of the
relation in Fig.\ref{fig:SM3e13_T}, using just the shape of the NFW
profile, coupled with entropy profiles similar to those seen in
simulations incorporating only gravity and shock heating. We therefore
conclude that there may be no need to invoke modification of the entropy
profiles by preheating
to explain this particular result, despite the initially surprising,
non-monotonic trend.

\section{Discussion}
\label{sec:discuss}
A substantial amount of theoretical and computational effort has been directed
towards the problem of similarity-breaking in groups and clusters,
especially over the past four years. We now compare the various models with
the results above, and with additional information from the companion
papers to this (Paper I and Paper II), to see how they fare.

\subsection{Cooling models}
A number of authors \citep{bry00,pea00,mua01,wuX02b,dave02} have
suggested that it may be unnecessary to invoke additional heating to
explain the entropy floor, since cooling will remove low entropy gas
from clusters. The counterintuive fact that cooling can have similar
effects to heating in this regard was first noted by \citet{kni97},
who found, on the basis of 1D hydrodynamical simulations of cluster
growth, that cooling acted to flatten the gas density profiles,
especially in low mass systems, and to steepen the $L$:$T$ relation,
but not sufficiently to match observations. More sophisticated 3D
simulations of cluster evolution by \citet{mua01} and \citet{dave02}
have shown that larger effects can be produced, depending upon the
spatial resolution of the simulations. \citeauthor{dave02} show that
their simulations have converged in this respect, although their
cooling function does not incorporate emission from metal lines, which
becomes very significant at $T<2$\keV.

Cooling achieves its effect of breaking the similarity between low and
high mass clusters, since, at a given gas density, the cooling time is
shorter at lower temperatures. Hence a larger fraction of the hot
baryons cool in groups, compared to richer clusters \citep{bry00}.
\citet{dave02} show that this is able to reproduce the cluster-group
$L$:$T$ relation quite well, except that the predicted steepening in
the relation falls at a slightly lower temperature than that
observed. They also compare the predicted gas entropies at 0.1\R200\
with the results of \citet{pon99} at $T<4$\keV, and find reasonable
consistency. However, agreement with the results from the present
study (Figs.~\ref{fig:S0.1_T} and \ref{fig:S0.1_T_grp}), which is
superior to that of \citet{pon99} in terms of sample size and
allowance for non-isothermality, is much less
good. \citeauthor{dave02} find that $S(0.1\R200)$ is raised in low
temperature systems, but converges with the self-similar trend through
hot clusters for systems with $T>3$\keV. In contrast, our results show
a relation which is flatter than the self-similar line across the full
temperature range. \citet{bry00} and \citet{wuX02b} produce results in
better agreement with our observations, but their analytical model is
based on an assumed variation in star formation efficiency with system
mass which is much stronger than that which we derive from the
subsample of our systems for which we have optical luminosities (Paper
II).

However, the most fundamental problem faced by cooling-only models, as
many of their proponents have acknowledged, is that cosmological simulations
without some form of effective `feedback' of energy into the baryonic
component as stars form, have a serious `overcooling' problem, which
has been recognised for many years. At high redshift, the Universe is dense,
and a large fraction of the baryons cool and form stars within small
collapsed haloes. For example, \citet{dave02} find that in the largest systems
in their simulation (with $T\sim 4$\keV) almost half of the baryons have
cooled out of the hot phase, whilst in small groups over 80\% of the
gas has cooled. Unless this cooled gas does not form stars, the very
high star formation efficiencies implied are far above those observed in
clusters (\cf\ discussion in Paper II) or inferred globally from the
K-band luminosity function of galaxies \citep{bal01}.
Furthermore, recent X-ray spectral observations from \XMM\ and \Chandra\
have made it clear that our understanding of cooling gas in the local
Universe, on scales ranging from clusters to elliptical galaxies,
is deficient \citep[see \eg][]{fab01}. It seems likely
that cooling rates have been seriously overestimated, for reasons
which are still under very active debate, but may be related to the
effects of feedback from AGN, which probably exist at the centres
of all cooling flows.

\subsection{Preheating models}
Since less energy is required to
raise the entropy of gas by a given amount when its density is lower,
it is energetically favourable to `preheat' gas before it is
concentrated into a cluster or group potential. A comparison of the
excess entropy required, with that potentially available from
supernova explosions associated with star formation
\citep[\eg][]{val99} shows that a high efficiency of heating of the
intergalactic gas is required even in the most favourable
circumstances -- \ie\ only a modest fraction of the supernova energy
must be radiated away. However, such high efficiencies have been
inferred on the basis of observations and modelling in the case of
starburst winds \citep{str00}, and so are not necessarily unreasonable
during the epoch of galaxy formation.

A large reservoir of energy at high redshift is potentially available
from the formation of massive black holes in galaxies
\citep{val99,wu00}, although it is at present unclear how much of this
energy can be coupled into heating of the IGM. Radiative heating from
AGN cannot achieve the required high entropies, but quasar outflows
may do so, although the physics and demographics involved are still
subject to considerable uncertainties \citep{nat02}.

Finally, cosmological simulations suggest that a large
fraction of the baryon content of the Universe is currently in the
form of what has been dubbed the `warm-hot intergalactic medium',
with an entropy of the order of 300\keV\cmsq\ \citep{cen99,dave01}.
\citet{dave01} show that this high entropy gas is primarily heated,
not by star-formation, but by the effects of shocks generated during
the collapse of gas into filaments. It seems unlikely, however, that
this mechanism can account for the excess entropy seen in the inner
regions of groups and clusters, since the entropy of the IGM generated
in this way declines steeply with redshift \citep{val02}, such that it
should be well below the observed floor value, at the epoch when the
gas in cluster cores was accreted. This is confirmed by the recent
simulations of \citet{bor02}, for example, which find baryon
distributions in clusters which are essentially self-similar, in the
absence of non-gravitational heating processes.

Pioneering efforts to explore the effects of preheating through
simulations or analytical treatments \citep{met94,nav95,cav97} have
been largely superseded by more recent work, and we concentrate here
on the latter.  Considerable insight can be gained by analytical
models in which preheated gas is accreted into clusters. The model of
Babul, Balogh and coworkers has reached its most advanced stage of
development in \citet{bab02}. This model assumes Bondi accretion
\citep{bon52} of preheated gas into a potential well represented by a
(fixed) NFW profile. This accretion is taken to be isentropic in low mass
systems, with introduction of a shock heated regime, motivated
by the entropy profiles seen in numerical simulations, for systems above
some critical mass. The level of preheating is tuned to achieve a
good match to the $L$:$T$ relation across a wide temperature range
(0.3--15\keV). Unfortunately, the entropy level required to achieve this
(330\keV\cmsq), is well above that actually observed in the inner
regions of groups. Moreover, the model predicts large isentropic
cores in cool systems. The IGM in groups with $M_{200}<8\times 10^{13}$\Msol\
(corresponding to $\sim 2$\keV, from our $M$:$T$ relation in Paper I) is
{\it entirely} isentropic in these models, in strong conflict with
our results (Fig.~\ref{fig:s_r}).

\citet{toz01} and \citet{dos02} have attempted to evaluate the effects
of shock heating on the preheated accreting gas. \citeauthor{dos02}
concentrate on the scaling properties of the gas entropy immediately
inside the accretion shock. They find that this is expected to scale
with mean system temperature according to $(1+T/T_0)$, where $T_0$ is
an adjustable parameter related to the initial adiabat on which the
preheated gas lies. By choosing the preheating entropy to be
120\keV\cmsq, corresponding to $T_0=2$\keV, they obtain a good match
to the $S(0.1\R200)$:$T$ plot of \citet{llo00}, though their curve is
rather too concave in comparison with the trend seen in
Fig.~\ref{fig:S0.1_T_grp}. In order to calculate the expected $L$:$T$
relation from their model, \citeauthor{dos02} make two key additional
assumptions: the IGM is assumed to be isothermal, and entropy
profiles, $S(r/\R200)$, are assumed to have the same shape in all
systems. Their resulting $L$:$T$ model steepens rather too gently to
match the group data (which show a very steep slope -- \eg\ $L\propto
T^{4.3}$ from \citet{hel00b}) entirely satisfactorily. The isothermal
assumption is in conflict with our data (Fig.~\ref{fig:t_r_grp}) and
with the results of cosmological simulations, which generally show a
decline in temperature by about a factor 3 from the inner regions to
the virial radius \citep{fre99}. A picture involving self-similar entropy
profiles, with a scale factor given by $(1+T/T_0)$, is in remarkably
good agreement with our results, as shown in Fig~\ref{fig:s_rDS}.
However, it should be noted that the \citet{dos02} model predicts only the
way in which entropy should scale just inside the shock -- the self-similarity 
in the radial distribution of entropy inside the shock is an {\it assumption},
rather than an output of their model.

A more comprehensive treatment is provided by \citet{toz01}, who
calculate entropy profiles based on solving the shock jump conditions
over typical accretion histories, for systems spanning a range of
final mass. They find that shock heating generates entropy profiles
with a characteristic logarithmic slope $S\propto r^{1.1}$, outside a
central isentropic core, which, for a given preheating level, occupies
a larger fraction of \R200\ in lower mass systems. The slope of 1.1
agrees well with our results (Fig.~\ref{fig:s_r}), and with numerical
simulations \citep[\eg][]{bor01}. However, as can be seen from 
Fig.~\ref{fig:s_r}, we do not see the larger isentropic cores in
$S(r/\R200)$ which appear to be a feature of all preheating models.

\citet{nat02} explore the energetics and evolutionary history of
preheating the IGM through outflows from radio loud, and broad absorption
line quasars. Their conclusion is that AGN could provide the
energy input required to account for the entropy floor. However, in
the context of our results, their most interesting result is 
that the specific energy injected by AGN into the gas,
is expected to be higher in low mass systems, for two reasons: the
incidence of powerful AGN is expected to be higher in smaller clusters,
and the fraction of their power expended in $P d\!V$ work is larger
in smaller systems. This appears to be contrary to our entropy
scaling results, shown in Figs.~\ref{fig:S0.1_T_grp} and \ref{fig:S500_T},
which show that at small radii the excess entropy is similar 
($\sim 100$\keV\cmsq) across a wide range of system masses, whilst at
larger radii it is actually highest in moderately rich ($T\sim 3$--4\keV)
clusters.

\subsection{Star formation models}
A natural development of cooling models, which can help to address the
overcooling problem, is the incorporation of feedback due to star
formation. In numerical studies, the successful implementation of such
a scheme is extremely challenging, and has been the Holy Grail of
those engaged in simulations of galaxy formation for some years.
\citet{voi01} introduced an interesting perspective on this issue, in the
context of similarity breaking in clusters, by noting that it makes
rather little difference quite how effective feedback is in heating
the IGM, since any gas within virialized systems which has low
entropy will have a short cooling time, and so is removed from its
location near the centre of a dark halo either by dropping out of the
hot phase, or being heated by star formation in its vicinity, such
that it escapes to large radii. In particular, they noted that the
entropy of gas with a cooling time equal to the age of the
Universe is $\sim 100$\keV\cmsq, in striking agreement with the
entropy floor reported in groups and clusters. Since the cooling
time scales as $T^{1/2}/n$ in systems with $T>2$\keV, where bremsstrahlung
dominates cooling, whilst $S=T/n^{2/3}$, it follows that a given
cooling time is achieved in gas with an entropy which scales
as
\begin{equation}
S_{\rm cool} \propto (t_{\rm cool} T)^{2/3} ,
\end{equation}
so that the entropy floor generated in this way is expected
to be higher in hotter systems \citep{voi01}. This feature may 
help to explain the flat trend in entropy with temperature apparent in 
Fig.~\ref{fig:S0.1_T_grp} -- note the similarity between
the dependence of $S_{\rm cool}$ on $T$, and the slope seen in
Figs.~\ref{fig:S0.1_T},\ref{fig:S0.1_T_grp} and \ref{fig:S500_T}.

This approach was developed further by \citet{voi02}, who constructed
models in which the entropy distribution is either truncated below
some critical entropy (corresponding to gas cooling out, or being
ejected as a result of vigorous heating), or shifted by the addition
of an entropy boost throughout (corresponding to preheating of the
entire IGM). The two different prescriptions for entropy modification
produce only subtle differences in observable properties: shifted
entropy models have rather flatter gas density profiles at large radii
in poor groups, and also slightly higher gas temperatures. These 
differences are not distinguishable with data of the quality 
used in the present study, and may well be too challenging even
for future studies with \XMM\ and \Chandra. One impressive feature
of these models is that they contain essentially no adjustable
parameters, since the entropy threshold is set by equating the cooling
time of gas to a Hubble time of 15~Gyr. However, the predicted $L$:$T$ 
relation fails to steepen sufficiently at low temperatures to pass through
the bulk of the galaxy group points, and the $M$:$T$ relation, whilst
steeper than the self-similar relation ($M \propto T^{1.5}$), is
less steep than that derived from the present sample in Paper~I.

One of the most recent numerical studies of cluster structure to
incorporate the effects of both cooling and feedback is the work of
\citet{bor01,bor02}. These simulations explore the effects of setting
an entropy floor through instantaneous preheating at high redshift,
and also through heating scaled to the expected supernova energy input
in overdense regions throughout the evolutionary history of the
Universe.  The effects of radiative cooling are also included in one
of the supernova heating runs. The main conclusions from this work,
are that $\ga 1$\keV\ per particle of energy input is required to give
a reasonable match to the observed steepening of the $L$:$T$ relation.
The authors point out that such a large energy boost appears to exceed
what is expected from supernova input alone, even if a high heating
effeciency of the IGM is assumed.

\citeauthor{bor02} find that a somewhat larger suppression of $L_X$ in
groups is achieved by injecting the same amount of energy
progressively, compared to global preheating of the IGM. Entropy
profiles generally agree well with the \citet{toz01} slope of 1.1, but
are flattened in low mass systems by the effects of energy injection,
as are gas density profiles. The gas temperature profile becomes more
centrally peaked in low mass haloes subject to additional heating. The
$M$:$T$ relation is found to be little affected by heating, either in
slope or normalization.  However, cooling is found to decrease the
normalization and steepen the slope of the relation, bringing it into
better agreement with observations. In comparison to these results, we
certainly observe flatter gas density profiles in cool systems
(Paper~I), and Fig.~\ref{fig:s_rDS} provides some tentative evidence
that the entropy slope may be rather shallower in lower temperature
($T<3$\keV) systems. In the case of the $M$:$T$ relation,
\citet{llo02} and \citet{voi02} show that systematic variations in the
concentration parameter of dark matter haloes with mass, may play a key
role in steepening its slope, as discussed in Paper~I.

%
%

\section{Conclusions and suggestions}
\label{sec:conc}
\subsection{Existing models}
Drawing the above results together, it seems that cooling-only models
cannot provide a viable explanation for the similarity-breaking
observed in clusters, unless one is willing to admit the presence
of very large quantities of baryonic dark matter, which would have
to dominate the baryon budget in groups.

Preheating models appear to have the generic property of generating
large isentropic cores in low mass systems, in conflict with our
observations. This seems to be an inescapable feature of simple
preheating models in which the IGM is raised uniformly
and `instantaneously' to a high adiabat, since such gas can only
be shocked when falling into potential wells deep enough for
its motion to be supersonic. More complex preheating models
may enable this problem to be circumvented, as we discuss below.

Models involving a mixture of cooling and star formation
(or possibly AGN heating) appear more natural and promising.
However, a number of features of our data appear to conflict with 
{\it all} models proposed to date:
\begin{enumerate}
\item The very large entropy excesses seen at large radii 
(Figs.~\ref{fig:S500_T} and \ref{fig:S500_Mnorm}) are a suprise.
Models tend to show entropy enhancement in the inner regions
of low mass systems, with a normal shock-generated entropy profile
re-establishing itself at larger radii.
\item Closely related to this, is the fact that the entropy
profiles appear to be approximately self-similar apart from
a normalization constant, and in particular, that larger 
isentropic cores are {\it not} seen in galaxy groups. In fact,
Fig.~\ref{fig:s_r} shows that the lowest temperature systems
actually have entropy profiles which appear to drop all the way
into the centre, unlike hotter systems. The fact that the
scaling suggested by \citet{dos02} brings our profiles into
good agreement (Fig.~\ref{fig:s_rDS}) is a puzzle, given that
this scaling is really not justified by their model as a scale
factor for entropy {\it profiles}.
\item The temperature profiles shown in Fig.~\ref{fig:t_r_grp}, are also
not quite what is expected from the models. Any mechanism which gives
an entropy boost should produce the strongest results in low mass systems.
Since a rise in entropy has to be coupled with the maintenance
of hydrostatic equilibrium, the natural consequence is a rise
in central temperature, coupled with a decrease in density.
The density drop is certainly observed, and there are indications
that the temperature has been raised in cool systems outside the core,
but the temperature profiles in groups seem to lack the expected
central cusp. This is related to the lack of an entropy core in groups
referred to in point (ii), above. 
\item The behaviour of the gas core radius, discussed in Paper~II, is
similarly unexpected. As a fraction of \R200, the gas core radius
is approximately constant at $\sim$10\% over most of our temperature
range, but falls dramatically, by an order of magnitude, in groups.
The only study to predict this sort of behaviour is that of
\citet{wuX02}, who show that fitting a convex profile which steepens
progressively, with a beta model, can lead to smaller fitted core
radii as the outer radius of detection shrinks. In Paper~I we reported
tests with fits to real data for two clusters, truncated at different
radii, which suggested that this effect is not dominating our fits. 
However, X-ray surface brightness profiles extending to larger
fractions of \R200\ for groups are required to definitively settle
this issue.
\end{enumerate}

\subsection{New possibilities}
What might these disagreements with the models be pointing towards?
\citet{fin02} argued that the large excess entropy at $R_{500}$ 
in groups indicates that the entropy profiles in cool systems are dominated
{\it throughout} by the effects of preheating. Two features of our
results make us uneasy about this conclusion. Firstly, these excesses
are now seen (Fig.~\ref{fig:S500_Mnorm}) to extend up to
moderately rich clusters, and to involve entropies several times
those expected in the IGM at the present epoch \citep{val99}. Secondly, the
slope of all our entropy profiles (Fig.~\ref{fig:s_r}) seem to scatter 
about the value predicted from shock heating. This is an uncomfortable
coincidence if shock heating plays no role in establishing them.
It is certainly possible to devise preheating scenarios which 
could reproduce any of our profiles, by varying the history of energy
injection, and the density gradient of the gas into which this
energy was injected, in the vicinity of systems of a given mass.
However, such a solution would require a large amount of fine
tuning. 

If, on the contrary, we wish to retain shock heating as the basic
mechanism responsible for generating the rising entropy profiles seen
outside the core in both clusters and groups, how can this be
achieved? The post-shock temperature is essentially determined
by the infall velocity of gas into a halo, which could be reduced
by pressure forces, but not increased above the free-fall value.
Hence, the only way to raise the whole entropy profile in low
mass systems, in the way implicit in Fig.~\ref{fig:s_r}, seems to be
to reduce the density of the pre-shock gas accreting into lower mass
systems, relative to that falling into rich clusters. In simple
spherical collapse models, the matter turning around and accreting into
haloes at a given epoch is expected to have a given
overdensity. However, in reality, cosmological simulations show us
that accretion is far from spherical, and that much of it takes place
along filaments. 

This change in topology does not in itself change the relationship
between the mass of virialized systems and the density of the gas
accreting into them, since structure formation is (almost)
self-similar.  Bigger systems have bigger filaments, and the mean
density of the accretion flow at a given epoch is still set by the
overdensity at turnaround, which is independent of system
mass. However, this situation changes if preheating of gas in the
filaments introduces an additional scale into the problem. In general,
the scale height of gas in filaments will be set by hydrostatic
equilibrium of the gas in the local gravitational potential of the
dark matter \citep[\eg][]{val02}. The temperature to which gas is
shock heated during the collapse of filaments will scale with their
mass, and hence the effect of any thermal input into the IGM from
galaxy formation or the growth of supermassive black holes will be
more prominent in the smaller filaments associated with lower mass
systems. Preheating will act to increase the scale height, and hence
to reduce the density of gas in filaments. Even in filamentary structures
which are still collapsing, preheating of the gas will increase its
pressure, and retard its collapse relative to that of the dark matter.

The calculations of \citet{toz01} show (albeit in the spherically
symmetric case) that even in the preheated case, the accretion
shock rapidly becomes strong once it is established, so that
the post-shock entropy of gas accreting into a cluster will scale
as $T_{200}/n_1^{2/3}$, where $n_1$ is the pre-shock
density of the gas, and $T_{200}$ is the virial temperature of the
cluster. Hence, if preheating reduces $n_1$ in lower mass
systems, relative to rich clusters, then the entropy will scale
sub-linearly with  $T_{200}$, which is what we observe.

It is worth noting some features of this model. Firstly, the effect we
invoke is intimately linked to the fact that the accreting gas is
largely confined to filaments, and hence will not be seen in
analytical and semi-analytical models, such as those of \citet{toz01}
and \citet{bab02}, which assume spherically symmetric accretion.  In
the spherical case, the gas has nowhere to expand to in response to
the initial preheating. Consequently any entropy rise due to preheating
serves to raise the temperature of the gas, rather than to lower its density.
In the lowest mass systems, this temperature rise can delay shock formation,
and hence result in an isentropic core, but in high mass systems (and at large
radii in low mass ones) it has
essentially no effect, since the post-shock temperature is just
$T=\frac{1}{3} \mu m_p v_i^2 $ \citep{toz01}, where $v_i$ is the velocity
at which gas, with mean particle mass $\mu m_p$, flows into the shock.
In contrast, if preheating serves primarily to reduce the density of the
pre-shock gas, then this will lead to a rise in post-shock entropy whatever
the shock strength, since (for a given shock Mach number) the post-shock 
density simply scales with $n_1$, so that the post-shock entropy
$S_2 \propto n_1^{-2/3}$.

Secondly, this mechanism is a very efficient way of raising the
entropy of intracluster gas for two reasons. As has been noted previously
\citep{pon99} a given injection of energy produces a larger rise in entropy
when the gas density is lower. Hence it is more effective to deposit energy
into low overdensity structures such as filaments, than into fully collapsed
systems like clusters. Moreover, if this injection places the gas onto
a higher adiabat by reducing its density, then the effects of the accretion
shock serve to raise it further. For example, \citet{dos02} show that
the ratio between the post- and pre-shock adiabats is well approximated by
\begin{equation}
S_2/S_1 = A (1+\frac{8 \mu m_p v_i^2}{51 k T_1}) ,
\end{equation}
where A is a constant of order unity. So, a rise in $S_1$ by some
factor, due to a density change (and hence with no change in $T_1$)
driven by preheating, will boost $S_2$ by the same factor. Hence the
shock has a multiplier effect, and if $S_2/S_1$ is large, a modest
rise (in absolute terms) in $S_1$, may result in a much larger rise in
$S_2$. Since the entropy will typically jump by an order of magnitude
in the accretion shock, as its temperature is raised from $\sim
10^6$~K to $10^7$-$10^8$~K, entropy excesses of the magnitude reported
here within clusters (i.e. 100-1000\keV\cmsq), might be generated from 
a rise in entropy of
only $\sim$10-100\keV\cmsq\ within the filaments which feed them.

Thirdly, if preheating acts primarily to reduce the density of the
pre-shock gas, rather than to increase its temperature, then the Mach
number of accretion shocks is not much reduced, compared to the
unheated case. The post-shock entropy profiles are therefore almost
entirely due to the evolution of the accretion shock, and
$S(r)\propto r^{1.1}$ profiles are expected, outside a small unshocked
central core, in good agreement with observations.

Whilst the viability of this model clearly cannot be investigated by
spherically symmetric analytical treatments, one might hope to find
evidence to test it from 3-dimensional simulations of cluster
formation.  Rather few studies have been published which incorporate
the effects of feedback into the IGM, together with a study of its
effects on the entropy profiles of clusters. The results from the work
of \citet{bor01,bor02} and \citet{mua02}, are not very encouraging.
In both these studies, the effects of heating of the IGM serve to
flatten the entropy profiles in clusters, with the effects being more
pronounced in lower mass systems. This accords better with
observations than do most 1-dimensional analytical models, insofar as
distinct isentropic cores are {\it not} generally present. However,
the progressive flattening of $S(r)$ towards lower mass systems does
not agree with our observed properties (Fig.~\ref{fig:s_r}). 

It is premature to conclude at this stage that the preheated filament
model can be rejected. Feedback is very poorly understood, and its
incorporation into numerical codes is notoriously difficult. In the
study of \citet{mua02}, feedback was included by simply raising the
temperature of all gas particles by 1.5\keV\ at a redshift of 4.
\citet{bor01} adopted the approach of imposing an entropy floor on the
gas in overdense ($\delta>5$) regions at a variety of redshifts from 1
to 5, whilst \citet{bor02} attempted to model the effects of energy
injection from galaxies in a more realistic way by using a
semi-analytic scheme to predict the supernova rate, and sharing the
resulting energy amongst gas particles in regions with
$\delta>50$. Unfortunately, the interplay of feedback and cooling is
extremely complex, and hard to model. Results tend to depend strongly
on spatial resolution, and current simulations (including those of
Borgani \etal\ and Muanwong \etal) generally fail to reproduce the
observed low cooled fraction of baryons. A wider exploration of the
effects of different feedback schemes, and a study of the density of
gas flowing into accretion shocks in simulations, should clarify
whether the mechanism proposed here is actually the dominant effect in
modifying the entropy of intracluster gas. At the same time, higher
quality observations from \XMM\ and \Chandra\ will provide more robust
determinations of observed entropy profiles in a wide range of
virialized systems. Initial results from \XMM\ for a small number
of groups \citep{mus03,pra03} confirm that no central isentropic
core is present in these systems.

\section*{Acknowledgments}

We are grateful to Ed Lloyd-Davies and Maxim Markevitch
for providing X-ray data and contributing to the original analysis,
and to Stefano Borgani, Arif Babul and Paulo Tozzi for helpful discussions
of the results presented here.
AJRS acknowledges financial support from the University of Birmingham. This
work made use of the Starlink computing facilities at Birmingham.


\bibliography{tjp_bibtex}
\label{lastpage}

\end{document}